\documentclass[12pt]{article}
\usepackage{epsf}

\def\simge{\mathrel{\rlap{\raise 0.511ex \hbox{$>$}}{\lower 0.511ex 
\hbox{$\sim$}}}}
\def\simle{\mathrel{\rlap{\raise 0.511ex \hbox{$<$}}{\lower 0.511ex 
\hbox{$\sim$}}}} 
\def\slash#1{\setbox0=\hbox{$#1$}\dimen0=\wd0                     
\setbox1=\hbox{/} \dimen1=\wd1 
\ifdim\dimen0>\dimen1                    \rlap{\hbox to 
\dimen0{\hfil/\hfil}} #1                        
\else                                       
      \rlap{\hbox to \dimen1{\hfil$#1$\hfil}}  
      /   \fi}                                         
\newcommand{\be}{\begin{equation}}
\newcommand{\ee}{\end{equation}}
\newcommand{\bea}{\begin{eqnarray}}
\newcommand{\eea}{\end{eqnarray}}
\newcommand{\ra}{\rightarrow}
\newcommand{\epse}{\varepsilon^{\prime}/\varepsilon}
\newcommand{\lsim}{\stackrel{<}{_\sim}}

\renewcommand{\Im}{{\rm Im}}
\renewcommand{\Re}{{\rm Re}}

\begin{document}

\begin{titlepage}
\begin{flushright}
\end{flushright}
\vskip 1. cm
\begin{center}
{\LARGE \bf SUSY contributions to the charge asymmetry 
in $K^\pm\ra\pi^\pm\ell^+\ell^-$ decays} 
\vskip 1.4cm 
\vskip 1.0  cm
{\large\bf A. Messina}\\
\vspace{0.8cm}
{\normalsize {\sl 
 Dip. di Fisica, Universit\`a di Roma ``La Sapienza'' \& 
INFN sezione di Roma,
\\ P.le A. Moro 2, I-00185 Rome, Italy. \\ \vspace{.25cm}
}
}
\vspace{.25cm}
\vskip1.cm
{\large\bf Abstract\\[10pt]} \parbox[t]{\textwidth}{ 
We analyse the contributions to the 
charge asymmetry in $K^\pm\to\pi^\pm\ell^+\ell^-$ decays
induced by gluino-exchange diagrams in the context of 
supersymmetric models with generic flavour couplings.
We show that sizeable deviations with respect to the 
Standard Model are possible only under special 
circumstances. Within this scenario we set an upper limit 
of about $10^{-3}$ for the relative charge asymmetry -- integrated 
for $M_{\ell^+\ell^-} > 2 M_\pi$ -- of 
both muon and electron channels. We also show that this 
limit is close to saturating a model-independent upper bound 
on the charge asymmetry derived from the present 
constraints on $\Gamma(K_L \to\pi^0 \ell^+ \ell^-)$.}
\end{center}
\vspace*{1.cm} 

\end{titlepage}

\paragraph{I.}
CP violation is one of most interesting and least 
known aspects of particle physics. 
The recent results of  KTeV \cite{KTeV} and  NA48 \cite{NA48} 
about $\epse$ have
unambiguously shown the existence of CP violation 
in $|\Delta S|=1$ transitions (the so-called {\em direct} CP violation)
and ruled out superweak scenarios.
The experimental measurements of $\epse$ are generally 
compatible with the theoretical expectations 
within the Standard Model (SM) \cite{epsp_th}. Nevertheless, 
the latter are affected by sizeable theoretical  
uncertainties, and it is difficult to constrain
non-standard effects. At the moment the 
theoretical uncertainties on  $\epse$ are so large that we cannot 
even exclude that this observable is completely dominated by 
new physics (NP) contributions. Given this situation, it is 
highly desirable to obtain new independent information about 
CP violation in $|\Delta S|=1$ transitions.

Among direct-CP-violating observables, particularly 
interesting are those accessible in rare decays.
In these processes the smallness of the 
SM contribution leads to identifying more easily a 
possible large NP effect, whereas the simplicity of the hadronic 
structure helps us to keep under control the theoretical 
uncertainties \cite{LP}. 
An observable that satisfies these requirements 
is the charge asymmetry in $K^\pm\to\pi^\pm\ell^+\ell\-$,
defined as 
\be
\Delta_{\ell}(s_0) \ =\  \frac{ \displaystyle\int_{q^2 > s_0} 
\left[ \frac{ {\rm d} \Gamma}{ {\rm d} q^2} 
       (K^+ \to\pi^+ \ell^+\ell^-)  
     - \frac{ {\rm d} \Gamma}{ {\rm d} q^2} 
       (K^- \to\pi^- \ell^+\ell^-) 
 \right]  }{ \displaystyle\int_{q^2 > s_0} 
 \left[ \frac{ {\rm d} \Gamma}{ {\rm d} q^2} 
       (K^+ \to\pi^+ \ell^+\ell^-)  
     + \frac{ {\rm d} \Gamma}{ {\rm d} q^2} 
       (K^- \to\pi^- \ell^+\ell^-) 
  \right] }~\ \ ,\label{intro}
\ee 
where $q^2 = M_{\ell^+\ell^-}^2$ is the dilepton invariant mass 
and $\ell = e$ or $\mu$.

This asymmetry is a pure direct-CP-violating observable. A non-zero 
$\Delta_{\ell}$ is generated by the interference between the absorptive 
contribution of the long-distance amplitude and a CP-violating 
phase of short-distance origin \cite{EPR}. As we shall discuss below, 
the kinematical cut on the dilepton invariant mass ($q^2 > s_0 $)
is a useful tool to maximize the CP-violating effect;  
indeed, the SM expectations of this asymmetry are 
given by \cite{DEIP}
\be
\left|\Delta_{e}(4 m_e^2 )\right|_{\rm SM} 
 \lsim 10^{-5} \qquad {\rm  and} \qquad 
\left| \Delta_{e,~\mu}(4 m_\pi^2 )\right|_{\rm SM} \lsim 10^{-4}~.
 \label{eq:SM}
\ee

The smallness of these figures leads to considering this
observable as a good probe of possible NP effects.
One of the most promising scenarios where 
sizeable non-standard CP-violating contributions 
could be generated is the supersymmetric (SUSY) 
extension of the SM with generic flavour
couplings and minimal particle content.
In particular, it has been recognized that 
in this context gluino-mediated penguin diagrams could 
naturally account for the observed value of $\epse$ \cite{MM}.
The purpose of this paper is to analyse the 
consequences that this scenario could 
have on $\Delta_{\ell}$. We shall therefore 
assume that the CP-violating phase of the 
$s\to d \ell^+\ell^-$ amplitude is entirely 
dominated by gluino-mediated penguin diagrams  
or, to be more specific, by the contributions 
of the dimension-5  chromomagnetic 
and electromagnetic dipole operators (CMO and EMO). 
The couplings of these operators, determined by 
the mismatch between quarks and squarks mass matrices,
appear also in $\varepsilon$ and 
$\varepsilon^\prime$. We will therefore extract 
the allowed range of these couplings from the  measured values of 
$\varepsilon$ and $\varepsilon^\prime$
in order to analyse the possible 
effects on~$\Delta_{\ell}$.

Our conclusions are that, in the general case,   SUSY 
effects are at most as large as the SM results 
in (\ref{eq:SM}) and thus not particularly 
interesting. Only assuming a cancellation between 
two independent SUSY contributions to $\varepsilon^\prime$,
or in a fine-tuned scenario, is it possible to reach 
higher values. In any case, within the minimal SUSY extension 
of SM considered here, $\Delta_{\ell}$ cannot exceed the 
$10^{-3}$ level (independently of the $q^2$ cut).
We will also show that this bound is very close to a 
model-independent limit on the charge asymmetry which was extracted, 
by means of isospin symmetry, from  the present 
constraints on $\Gamma(K_L \to\pi^0 \ell^+ \ell^-)$.

This paper is organized as follows: we shall first discuss 
the amplitude decomposition of $K^{\pm} \to\pi^{\pm} \ell^+ \ell^-$ decays
and the generic expression of the charge asymmetry; then we will 
introduce the effective Hamiltonian describing the SUSY short-distance 
contributions; finally we shall discuss the phenomenological 
constraints on the SUSY phases and the corresponding 
bounds for the charge asymmetry.

\paragraph{II.}
The charge asymmetry is produced by interference between 
the CP-conserving strong phase and the CP-violating weak phase
of the decay amplitude. The latter is generated by the 
exchange of heavy particles (short-distance), the former is 
due to the $K \to 3 \pi$ intermediate state and thus belongs 
to the long-distance part of the amplitude 
($K^\pm \to 3 \pi \to \pi^\pm\ell^+\ell^-$).
As shown in \cite{EPR}, the decay width of $K^\pm\to\pi^\pm\ell^+\ell^-$ 
is largely dominated by long-distance effects and, as long as 
CP violation is not considered, short-distance contributions
can be neglected. 

In the long-distance part of the $K^\pm\to\pi^\pm\ell^+\ell^-$  
amplitude the two leptons are always produced by a virtual 
photon ($K^\pm\to\pi\gamma^*\to\pi^\pm\ell^+\ell^-$).
Since we are interested in the interference between short- and long-distance 
terms, all short-distance contributions where the 
two leptons are not in a vector state can be neglected. 
We can therefore parametrize the decay amplitude in terms 
of a single vector form factor, $W(z)$, defined as \cite{DEIP}
\be
{\cal A}\left( K^\pm(k)\to\pi^\pm(p)\ell^+(l^+)\ell\-(l^-) \right)
=-\frac{e^2G_F}{(4\pi)^2}W(z)(k+p)^{\mu}\bar{u}(l^+) \gamma_{ \mu}v(l^-)~,
\label{ampiezza finale k3pi}\\
\ee
where $q=k-p$ and $z=q^2/M_K^2$. 
Integrating  this amplitude over the phase space leads to 
\be
\frac{d\Gamma}{dz}=\frac{\alpha^2G_F^2M_K^5}{12\pi(4\pi)^4}
\rho(z)|W(z)|^2,\label{fasi}
\ee
where  $4r_\ell^2\le z\le (1-r_\pi)^2$, $r_i=m_i/M_K$ 
and $\rho(z)=\sqrt{1-4r^2_e/z}(1+2r^2_e/z)\lambda^{3/2}(1,z,r^2_{\pi})$.

Referring to the discussion in Ref.~\cite{DEIP}, 
we decompose the form factor as
\be
W(z)=W^{pol}(z)+\frac{1}{G_FM_K^2}W^{\pi\pi}(z)~.\label{ww}
\ee
Here $W^{\pi\pi}(z)$ denotes a non-analytic contribution,
largely dominated by the dipion intermediate state,  as a consequence
of the small $q^2$ of the lepton pair, which  
exhibits a branch cut along the real axis starting at 
$z=4r^2_{\pi}$.  The polynomial term 
$W^{pol}(z)$ includes both long- and short-distance contributions.
The latter are completely dominant in the real part of
$W^{pol}(z)$, which can be fitted from experiments.
Parametrizing the polynomial term as 
$W^{pol}(z)= G_F M_K^2 (a+b z)$,
consistently with a chiral expansion at the 
next-to-leading order \cite{DEIP}, the 
values of $a$ and $b$ fitted by the 
BNL E865 Collaboration are reported in Table~\ref{ab}. 
On the other hand, the imaginary part of $W^{pol}$ can  
 be produced only by the direct CP-violating 
weak phase of short-distance origin. 

\begin{table}[t]
\begin{center}
\begin{tabular}{||c|c||c|c||}
\hline
$ \Re(a)$ &$ \Re(b)$  &  $ \alpha$ &$ \beta$\\
\hline 
$(-0.587\pm 0.010)$ & $(-0.655\pm0.044)$ 
& $(-20.6\pm 0.5)\times 10^{-8}$ & $(-2.4\pm1.2)\times 10^{-8}$\\
\hline
\end{tabular}
\caption{\label{ab}\sl
Experimental values of the parameters determining 
$W^{pol}$ \cite{Appel} and $W^{\pi\pi}(z)$
\cite{DEIP,Kambor:ah} }
\end{center}
\end{table}

The function  $W^{\pi\pi}(z)$  has already been computed in the literature 
\cite{EPR, DEIP} and we report it here for completeness:
\be
W^{\pi\pi}(z) =  \frac{1}{r_{\pi}^{2}}\left[\alpha+\beta\frac{z-z_0}
{r_{\pi}^{2}}
\right]\Phi(z)\chi(z),\label{wpipi}
\ee
where the values of $\alpha$ and $\beta$ (fitted from $K\to 3 \pi$ decays)
are shown in Table~\ref{ab}, $z_0=1/3+r_{\pi}^{2}$, $\Phi(z)=1+z/r_V^2$ 
($r^2_V \simeq 2.5$) 
and  $\chi(z)$ is defined by\footnote{~$\chi$ satisfies the  relation: 
$\chi(0)= 0$; this condition follows from the 
possibility to include $\chi(0)$  into a redefinition of the polynomial form
factor.}
\be
\chi(z)= \frac{4}{9} - \frac{4}{3} \frac{r_{ \pi}^{2}}{z} - \frac{1}{3}
\left (1 - \frac{4r_{\pi}^{2}}{z} \right ) G(z/r_{\pi}^{2})~,\label{wpipifin}
\ee
\be
G(z/r_{ \pi}^{2}) = \left\{
\begin{array}[l]{cc}
\sqrt{4r_{\pi}^{2}/z-1} \arcsin( \sqrt{z/4r_{\pi}^{2}}) & z \leq 4r_{\pi}^{2}~,\\
- \frac{1}{2} \sqrt{1-4r_{\pi}^{2}/z} \left ( \log \frac{1- 
\sqrt{1-4r_{\pi}^{2}/z}}{1+ \sqrt{1-4r_{\pi}^{2}/z}}+ i\pi\right ) & 
z \geq 4r_{\pi}^{2}~. \\
\end{array}  \right.\label{G}
\ee

The interference between the CP-violating phase in 
$W^{pol}$ and the absorptive contribution to $W(z)^{\pi\pi}$
leads to the following difference between the widths of the 
charge-conjugated modes
\be
\Gamma^+_{\ell}-\Gamma^-_{\ell}=\frac{\alpha^2G_F^2M_K^5}{12\pi(4\pi)^4}
\int_{4r_{l}^2}^{(1-r_{\pi})^2}4 \rho(z)\left[\Im(W(z)^{pol})\cdot\Im
(W(z)^{\pi\pi})\right]dz~.\label{g+-}
\ee
Obviously, in order to obtain an adimensional asymmetry 
we should normalize the width difference to the sum of the widths. 
However, we stress here the strong dependence of this observable
from the dilepton invariant mass: owing to the kinematical threshold in the 
absorptive part, only when the two-pion intermediate state can be on shell 
is the asymmetry  different from zero. 
This observation suggests the construction of integrated 
asymmetry setting a cut on $q^2$ above  $4 m_{\pi}^2$,
i.e. setting $s_0 = 4 m^2_{\pi}$ in Eq.~(\ref{intro}).
Since electron and muon channels have a very similar
phase space for $q^2 \ge 4m^2_{\pi}$, with such a cut 
the normalized asymmetry turns out to be 
very similar in the two  modes. 
On the other hand, since most of the available phase space 
for the electron channel is below this cut, from the 
experimental point of view the muon channel appears a
better candidate for the study of this asymmetry. 

\paragraph{III.}
As discussed in the previous section, we look for short-distance
contributions to the $s\to d \ell^+\ell^-$ amplitude,
with the lepton pair in a vector state and possibly  
with a sizeable new weak phase. As shown in Ref.~\cite{Buras:1999da}, 
in the presence of non-minimal flavour mixing, the largest SUSY contributions 
to this type of amplitude are produced by the dimension-5 CMO and EMO. 
All the other contributions are in fact naturally
suppressed once the bounds from other processes
are taken into account (barring accidental cancellations). 
The structure of the effective Hamiltonian necessary to describe these 
contributions reads
\be
{\mathcal{H}}_{eff}=C^{+}_{\gamma}Q^{+}_{\gamma}+C^{-}_{\gamma}Q^{-}_{\gamma}
+ C^{+}_gQ^{+}_g+C^{-}_gQ^{-}_g + {\rm h.c.}~,\label{H}
\ee
with the operators expressed in the following basis
\be
Q^{ \pm}_{\gamma}= \frac{e Q_{\tilde{d}}}{16 \pi^{2}} \left( \bar{s}_{L} 
\sigma^{ \mu \nu}
F_{ \mu \nu}d_{R} \pm \bar{s}_{R} \sigma^{ \mu \nu}F_{ \mu \nu}d_{L}  
\right ),
\label{qt}
\ee
\be
Q^{ \pm}_g= \frac{g}{16 \pi^{2}} \left( \bar{s}_{L} 
\sigma^{ \mu \nu}
t^aG^a_{ \mu \nu}d_{R} \pm \bar{s}_{R} \sigma^{ \mu \nu}t^aG^a_{ \mu \nu}d_{L}
 \right )~.
\label{qt2}
\ee
Note that in this basis the  operators have well defined properties 
under parity transformations: $Q_i^+$ induce
parity-conserving transitions  and $Q_i^-$ parity-violating ones.

The Wilson coefficients of these operators induced by gluino
 exchange  are given by \cite{Gabbiani:1996hi, Buras:1999da}
\be
C^{\pm}_{\gamma}(m_{\tilde{g}})=
F(x)\frac{\pi \alpha_{s}(m_{\tilde{g}})}{m_{\tilde{g}}}
\left[(\delta_{LR})_{21} \pm
(\delta_{LR})_{12}^{*}\right],\label{ct}
\ee
\be
C^{\pm}_g(m_{\tilde{g}})=
G(x)\frac{\pi \alpha_{s}(m_{\tilde{g}})}{m_{\tilde{g}}}
\left[(\delta_{LR})_{21} \pm
(\delta_{LR})_{12}^{*}\right].\label{ct2}
\ee
Here $(\delta_{LR})_{ij}=(M^2)_{i_{L}j_{R}}/m^2_{\tilde{q}}$ denotes the 
off-diagonal entries of the  down-type matrix in the super-CKM basis, and
$x=m^2_{\tilde{g}}/m^2_{\tilde{q}}$ the ratio of gluino and 
(average) squark masses
 squared. The explicit expression of the loop functions 
$F(x)$, $G(x)$ can be found in \cite{Buras:1999da}.
We have also considered  the CMO, even if it does not participate 
at  tree level, because of the large mixing between
CMO and EMO induced by QCD interactions at the one-loop level. 
Taking into account the $2\times 2$ 
anomalous-dimension matrix of these operators \cite{Buras:1999da},
computed at lowest order, the Wilson coefficients evolved down 
to charm scales  reads 
\bea
C_{\gamma}^{\pm}(m_c)& =& \eta^2[C_{\gamma}(m_{\tilde{g}}) +8(1-\eta^{-1})
C^{\pm}_{g}(m_{\tilde{g}})], \label{eq:Cgamma1}\\
C^{\pm}_g(m_c)&=&C^{\pm}_{g}(m_{\tilde{g}}),
\eea
where
\be
\eta =\left(\frac{\alpha_s(m_{\tilde{g}})}{\alpha_s(m_t)}\right)^{\frac{2}{21}}
\left(\frac{\alpha_s(m_t)}{\alpha_s(m_b)}\right)^{\frac{2}{23}}
\left(\frac{\alpha_s(m_b)}{\alpha_s(m_c)}\right)^{\frac{2}{25}}
= 0.89\left( \frac{\alpha_s(m_{\tilde{g}})}{\alpha_s(500~\mathrm{GeV})}
\right)^{\frac{2}{21}}~.
\ee
Starting from Eq.~(\ref{eq:Cgamma1}), we find it convenient to rewrite 
the Wilson coefficient of the EMO as 
\be
C_{\gamma}^{\pm}(m_c) = \frac{\pi \alpha_s(m_{\tilde{g}})}{m_{\tilde{g}}}
Y(x) \delta^{\pm},\label{Cgamma}
\ee
where
\be
Y(x)=
G(x)\eta^2\left[\frac{F(x)}{G(x)}+8(1-\eta^{-1})\right],\label{Lambda}
\ee
\be
\delta^{\pm}=(\delta_{LR})_{sd}\pm (\delta_{LR})_{ds}^*~.
\label{eq:deltapm}
\ee 

In order to identify the SUSY contribution to the polynomial form
factor, we need to evaluate the EMO matrix element 
between $K^+$ and $\pi^+$ external states:
\be
\langle\pi^{+}|Q^{+}_{\gamma}
|K^{+}\rangle= 2i\frac{eQ_{\tilde{d}}}{16\pi^{2}}
\frac{B_{T}}{M_{K}} p_{\mu}(\pi^{+})p_{\nu}(K^{+})F^{\mu \nu}.
\label{operatorem}
\ee
As is usually done in the literature, we have expressed the hadronic 
matrix element in terms of a suitable $B_T$ parameter \cite{Buras:1999da,CIP}, 
expected to be ${\cal O}(1)$, which encodes  
the non-perturbative dynamics.
This parameter has recently been computed  
on the lattice \cite{Becirevic:2000zi}, 
confirming the estimate $B_T \approx 1$ made in Ref.~\cite{CIP}. 
We shall leave it as a parameter through all the paper, but 
for completeness we report here the recent lattice result 
\be
B_T(\mu=2~\mathrm{GeV})=1.21\pm 0.09\pm0.04^{+0.07}_{-0.00} \qquad
\protect\cite{Becirevic:2000zi}.
\ee

Now using Eqs.~(\ref{Cgamma})--(\ref{operatorem}) we can write 
the full matrix element of the Hamiltonian (\ref{H}) relevant 
to $K^{\pm}\to \pi^{\pm} \ell^+ \ell^-$ decays
\be
\langle \pi^{\pm} \ell^+ \ell^-|{\mathcal H}_{eff}|
K^{\pm}\rangle =
\alpha\alpha_s\frac{Q_{\tilde{d}}}{4}\left[\frac{B_T}{M_K
m_{\tilde{g}}}Y(x) \delta^+\right]
\times(k+p)^{\mu}\bar{u}(l^+)\gamma_{\mu}v(l^-).\label{htrastati}
\ee
Then, according to the definition of $W(z)$ in 
Eq.~(\ref{ampiezza finale k3pi}), 
we obtain: 
\bea
\Im(W^{pol}_{\mathrm{SUSY}}) &=& \pi \alpha_s(m_{\tilde{g}})
\frac{Q_d}{G_F}\frac{B_T}{M_K}
\frac{Y(x)}{\tilde{m}}\Im(\delta^+)~, \\ &=& 15.2\times
\left(
\frac{Y(x)}{Y_0(1)}\right)
\left(\frac{500~\mathrm{GeV}}{\tilde{m}}\right)
\left(\frac{\alpha_s(m_{\tilde{g}})}{\alpha_s(500~\mathrm{GeV})}
\right)
B_T\Im(\delta^+)~,
\eea
where $Y_0(1)=Y(x=1;~m_{\tilde{g}}=500~\mathrm{GeV})=0.39$. 
Inserting this result in Eq.~(\ref{g+-}) we can finally write: 
\be
\left|\Delta_e(4m_{\pi}^2)\right|=(1.0\pm 0.1)
\left[
\left(\frac{Y(x)}{Y_0(1)}\right)
\left(\frac{500~\mathrm{GeV}}{\tilde{m}}\right)
\left(\frac{\alpha_s(m_{\tilde{g}})}{\alpha_s(500~\mathrm{GeV})}\right)
\right]
\times B_T|\Im(\delta^+)|,\label{DMU}
\ee
where the error includes the uncertainty in the experimental parameters.
\paragraph{IV.}
Before starting a numerical analysis of the SUSY contribution 
to the charge asymmetry, we discuss here a more general 
upper bound on $|\Im(W^{pol})|$ (and thus on the charge asymmetry),
which can be extracted from the experimental upper bound on
$\Gamma(K_L\to\pi^\pm \ell^+ \ell^-)$. 

Isospin symmetry relates in a model-independent way the 
short-distance components of $K^\pm\to\pi^\pm \ell^+ \ell^-$ and 
$K_L\to\pi^\pm \ell^+ \ell^-$ amplitudes. Indeed, in both cases,  
the hadronic current is necessarily a $\Delta I=1/2$ operator of 
the type $\bar s \Gamma d$ (or $\bar d \Gamma s$). In the $K_L$ case, 
the approximate CP-odd combination of $K^0$ and $\bar K^0$ states 
select the imaginary part of this amplitude, which can therefore
be constrained by using the experimental bounds 
on $\Gamma(K_L\to\pi^\pm \ell^+ \ell^-)$.
Assuming that long-distance contributions to 
$K_L\to\pi^\pm \ell^+ \ell^-$ are negligible (which is certainly 
a good approximation if $\Gamma(K_L\to\pi^\pm \ell^+ \ell^-)$ saturates 
its current experimental bound \cite{LP}), we can write 
\be
\left| \Im(W^{pol}) \right| \leq \left[
\frac{12\pi(4\pi)^4}{\alpha G_F^2 M_K^5}
\frac{1}{\tau_L\int\rho(z) dz}
\mathrm{BR}(K_L\to\pi^0\ell^+\ell^-)\right]^{\frac{1}{2}},\label{Wlimit}
\ee
where $\tau_L$ is the mean lifetime of the $K_L$ and the integral on $\rho(z)$
extends to the whole phase space. 
Using this relation we obtain the following  
model-independent upper limits on the charge asymmetries of 
electron and muon modes\footnote{We report
separately the bounds on electron and muon modes, obtained by the
corresponding $K_L\to\pi^0 \ell^+ \ell^-$ decay, in order to
take into account possible violations of lepton universality.
}:
\bea
\left| \Delta_e (4m_{\pi}^2) \right| &\leq& 43
\times\sqrt{\mathrm{BR}(K_L\to\pi^0 e^+ e^-)}   \leq 1.0 \times 10^{-3}~, \\
\left| \Delta_\mu (4m_{\pi}^2) \right| &\leq& 82
\times\sqrt{\mathrm{BR}(K_L\to\pi^0 \mu^+ \mu^-)}   \leq 1.6  \times 10^{-3}~,
\eea
where the numerical values have been obtained using the  
experimental bounds BR$(K_L\to\pi^0e^+e^-)< 5.8 \times 10^{-10}$
and BR$(K_L\to\pi^0\mu^+\mu^-)<3.8 \times 10^{-10}$ obtained by the 
KTeV Collaboration \cite{KTeV_rari}.

In Table~\ref{asimm} we compare the model-independent upper bounds 
for the charge asymmetries with SM and  MSSM expectations,  
expressed in terms of the imaginary parts of 
the CKM factor $\lambda_t = V_{ts}^* V_{td}$ 
and the SUSY parameter $\delta^+$, respectively.
Since $\Im(\lambda_t) \approx 10^{-4}$, it is clear that 
the SM could only account for a few per cent of the general bounds. 
From this perspective it is particularly interesting to understand 
how large  the SUSY contribution could be.
 
\smallskip
\begin{table}[t]
\begin{center}
\begin{tabular}{|c|c|c|c|}
\hline
$ ~$ & SM  & SUSY & Model-Independent\\
\hline\hline
$|\Delta_e(4m^2_e)|$ & $0.07\times |\Im(\lambda_t)|$ & 
$0.13\times |\Im(\delta^+)|$ &$1.3\times10^{-4}$
\\
\hline
$|\Delta_e(4m^2_{\pi})|$ &$0.53\times |\Im(\lambda_t)|$ & 
$1.0\times |\Im(\delta^+)|$ &$1.0\times10^{-3}$
\\
\hline
$|\Delta_{\mu}(4m_{\mu}^2)|$ &$0.21\times |\Im(\lambda_t)|$ & 
$0.40\times |\Im(\delta^+)|$ & $6.4\times10^{-4}$
\\
\hline
$|\Delta_{\mu}(4m_{\pi}^2)|$ & $0.54\times|\Im(\lambda_t)|$&
$1.0\times |\Im(\delta^+)|$ & $1.6\times10^{-3}$
\\
\hline
\end{tabular}
\caption{\label{asimm}\sl{ Summary of the results for both the electron 
and muon channel asymmetry, evaluated in the SM \cite{DEIP} 
 and in the MSSM, with and 
without the kinematical cut on the dilepton square mass.}}
\end{center}
\end{table}
\smallskip

The bounds on $\Im (\delta^+)$ [and more in 
general on the $\Im(\delta_{LR})_{ij}$'s]
have been widely discussed in the recent literature (for  extensive 
discussions, see Ref.~\cite{D'Ambrosio:1999jh}
and references therein). In general there are two  classes of  
bounds: those coming from $|\varepsilon|$ and BR$(K_L \to \pi^0 e^+ e^-)$ 
(direct bounds) and those extracted from  $\Re(\epse)$
(indirect bounds). In the first case $\Im(\delta^+)$ is directly 
involved, whereas in the second case -- dealing with parity-violating 
transitions -- 
only $\Im(\delta^-)$ is directly involved.
As clearly shown by Eq.~(\ref{eq:deltapm}), $\delta^+$ and $\delta^-$ are 
naturally related: if we assume these two quantities to be 
of the same order of magnitude, then the measurement of 
$\Re(\epse)$ leads to a  bound 
on $|\Im(\delta^+)|$ of about $10^{-5}$ \cite{MM}. In this case the SUSY 
contribution to the charge asymmetries could be at most of the same 
order as  the SM one. 
On the other hand, if we allow a fine-tuned scenario where 
$|\Im (\delta^+)| \gg |\Im (\delta^-)|$, then only the (weaker) direct 
limits from $|\varepsilon|$ and BR$(K_L \to \pi^0 e^+ e^-)$ have to be satisfied. 
The constraints from BR$(K_L \to \pi^0 e^+ e^-)$ are  
equivalent to the model-independent bounds reported in Table~\ref{asimm}
and thus they certainly allow for sizeable deviations with respect to the SM.
In the case of $|\varepsilon|$ there is a considerable uncertainty, induced by 
hadronic matrix elements: employing optimistic 
values for the latter (or allowing cancellations with other SUSY effects)
it is possible to obtain bounds on $\Im (\delta^+)$
above $10^{-4}$ \cite{D'Ambrosio:1999jh}. 
Under these special circumstances the SUSY contribution to the charge 
asymmetry could become substantially larger than the SM one,
reaching values close to the model-independent bounds.

\begin{figure}[t]
\begin{center}
\epsfxsize13.0cm\epsffile{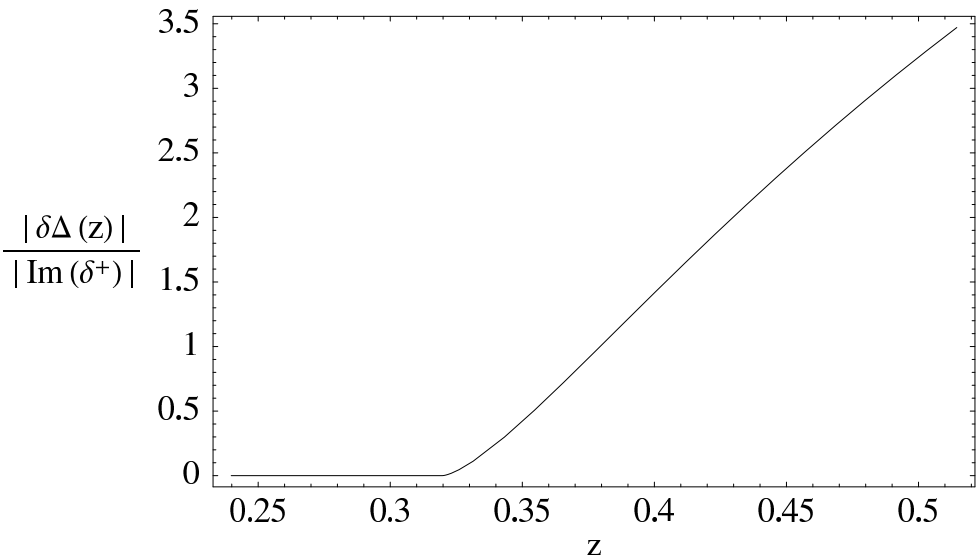} 
\caption{\label{adif}{\sl Electron differential asymmetry 
 as a function of $z$.}}
\end{center}
\end{figure}
Before concluding, we point out that an observable particularly 
useful from the experimental point is the differential asymmetry,
defined by: 
\be
\delta \Delta_\ell(z)= \frac{d\Gamma^+_{\ell}(z)/dz-d\Gamma^-
_{\ell}(z)/dz}{d\Gamma^+_{\ell}(z)/dz+d\Gamma^-_{\ell}(z)/dz}~.
\label{simmdiff}
\ee
This observable could be particularly useful in an experimental set up 
where it is difficult to obtain a precise flux normalization 
and thus precise width measurements. In Figure~\ref{adif} we plot 
$\delta \Delta_\ell(z)$, normalized to the SUSY phase $\Im(\delta^+)$, 
as a function of $z$.  

\paragraph{V.}
In this letter we have analysed the charge asymmetry in 
$K^{\pm}\to\pi^{\pm}\ell^+\ell^-$ decays in the framework 
of supersymmetric models with minimal particle content  
and generic flavour couplings. We have shown that in general the 
expectations are  of the same order of magnitude
as in the SM case. Nevertheless, under special circumstances it is possible 
to relax the indirect constraints on the SUSY CP-violating phases
and obtain results in the range  
$10^{-4} < \Delta_{\ell} (4m_\pi^2) \lsim 10^{-3}$, or above 
the SM expectation. 
We have also shown that under general assumptions $\Delta_{\ell}(4 m_\pi^2)$ 
cannot exceed the $10^{-3}$ level. This model-independent bound  
is derived, using isospin invariance, from the 
experimental constraints on $\Gamma(K_L\to\pi^0\ell^+\ell^-)$.

The first measurement of the charge asymmetry in the muon 
mode has recently been announced  by the HyperCP collaboration \cite{HyperCP}:
\be
\Delta_\mu(4 m_\mu^2) = -0.02 \pm 0.11~({\rm stat})~ \pm 0.04~({\rm syst})~.
\ee
Unfortunately at the moment the sensitivity is very far  from 
the interesting region, still well above the model-independent bound. 
In view of possible improvements of this 
measure, we can summarize as follows the possible outcome of 
a non-vanishing result:

\begin{itemize}

     \item    $\Delta_{\ell}(4m_\pi^2) \lsim 10^{-4}$~:  
              we could conclude that the charged kaon system is well described
              by the SM. Moreover, we could put a new strong constraint
              on the SUSY phase $\Im(\delta^+)$.

     \item    $10^{-4} < \Delta_{\ell}(4m_\pi^2) \lsim 10^{-3}$~: 
              we would have a clear signal of 
              new physics, compatible with the results obtained
              within the Minimal Supersymmetric Model with generic flavour couplings.

\end{itemize}

\subsection*{Acknowledgements}
I wish to thank G. Isidori for interesting me to this subject
 and for  useful discussions. I would also to acknowledge the TH Division
of CERN where this work was completed.

\end{document}